\documentclass[twocolumn,amsmath,aps]{revtex4}

\usepackage{amssymb,mathrsfs,bm,color,times}
\usepackage{graphicx,color}
\usepackage{CJK}
\usepackage{bm}
\usepackage[hypertex]{hyperref}
\usepackage{amsmath}

\newcommand{\be}{\begin{equation}}
\newcommand{\ee}{\end{equation}}
\newcommand{\bea}{\begin{eqnarray}}
\newcommand{\eea}{\end{eqnarray}}
\newcommand{\bsube}{\begin{subequations}}
\newcommand{\esube}{\end{subequations}}

\newcommand{\Eq}[1]{Eq.\,(\ref{#1})}

\newcommand{\la}{\langle}
\newcommand{\ra}{\rangle}

\newcommand{\beq}{\begin{equation}}
\newcommand{\eeq}{\end{equation}}
\newcommand{\beqn}{\begin{eqnarray}}
\newcommand{\eeqn}{\end{eqnarray}}

\newcommand{\nl}{\nonumber \\}

\newcommand{\bsub}{\begin{subequations}}
\newcommand{\esub}{\end{subequations}}

\begin{document}
\begin{CJK*}{GBK}{Song}

\title{Quality analysis for precision metrology based on joint weak measurements
without discarding readout data}
\author{Lupei Qin}
\email{qinlupei@tju.edu.cn}
\affiliation{Center for Joint Quantum Studies and Department of Physics,
School of Science, \\ Tianjin University, Tianjin 300072, China}

\author{Luting Xu}
\email{xuluting@tju.edu.cn}
\affiliation{Center for Joint Quantum Studies and Department of Physics,
School of Science, \\ Tianjin University, Tianjin 300072, China}

\author{Xin-Qi Li}
\email{xinqi.li@tju.edu.cn}
\affiliation{Center for Joint Quantum Studies and Department of Physics,
School of Science, \\ Tianjin University, Tianjin 300072, China}

\date{\today}

\begin{abstract}
We present a theoretical analysis for the metrology quality of
joint weak measurements (JWM),
in close comparison with the weak-value-amplification (WVA) technique.
We point out that the difference probability function
employed in the JWM scheme cannot be used to calculate
the uncertainty variance and Fisher information (FI).
In order to carry out the metrological precision, we reformulate
the problem in terms of difference-combined stochastic variables,
which makes all calculations well defined.
We reveal that, in general,
the metrological precision of the JWM scheme
cannot reach that indicated by the total FI,
despite that all the readouts are collected without discarding.
We also analyze the effect of technical noise, showing that
the technical noise cannot be removed by the subtracting procedure,
which yet can be utilized to outperform the conventional measurement,
when considering the imaginary WV measurement.
\end{abstract}


\maketitle

\section{Introduction}

Based on the concept of quantum weak values (WVs) \cite{AAV88,AV90}
proposed by Aharonov, Albert and Vaidman (AAV),
a novel precision metrology scheme termed as weak-value amplification (WVA)
has been developed and received considerable attentions
\cite{Kwi08,How09a,How09b,How10a,How10b,Guo13,Sim10,Ste11,Nish12,Ked12,
Jor14,Li20,Bru15,Bru16,How17,Lun17,Zen19,ZLJ20,Jor13,Sim15,Jor21}.
The WVA technique can lead to experimental sensitivity
beyond the detector's resolution.
It allowed to use high power lasers with low power detectors
while maintaining the optimal signal-to-noise ratio,
and obtained the ultimate limit
in deflection measurement with a large beam radius.
The WVA technique can outperform conventional measurement in the presence
of detector saturation and technical imperfections \cite{Lun17,Zen19,ZLJ20}.
Also, it has been pointed out that the WVA technique can reduce
the technical noise in some circumstances
\cite{Kwi08,How09a,How09b,Sim10,Ste11,Nish12,Ked12,Jor14,Li20},
and can even utilize it to outperform standard measurement
by several orders of magnitude by means of the imaginary weak-value measurements
\cite{Sim10,Ste11,Ked12,Jor14,Li20}.

The WVA technique involves an essential procedure termed as post-selection,
which discards a large portion of output data.
However, it was proved that WVA technique can
put almost all of the Fisher information
about the parameter under estimation
into the small portion of the remained data
and show how this fact gives technical advantages \cite{Ked12,Jor14,Li20}.
Since such result is possible under an almost orthogonal
pre- and post-selection procedure,
which leads to ultra-small probability of post-selection,
the WVA technique has caused controversial debates in literature
\cite{Nish12,Ked12,Jor14,Li20,Tana13,FC14a,Kne14,ZLJ15,Aha15,Li16,FC14b}.

Taking a different strategy,
the possibility of inducing anomalous amplification
was proposed without discarding readout data, but instead using all
the post-selection accepted (PSA) and post-selection rejected (PSR) data
\cite{Bru13,ABWV16,ABWV17a,ABWV17b,Zeng16}.
This proposal was referred to as joint-weak-measurement (JWM) scheme,
since, in the presence of the post-selection classification,
the system state and meter's wavefunction are {\it jointly} measured.
However, being different from the WVA technique,
where the intensity of the PSA data (in the so-called dark port) is very weak,
in the JWM scheme,
the intensities of the PSA and PSR data can be set {\it almost equal}
and the difference between them reveals anomalous amplification \cite{Bru13,ABWV16}.
For this reason, this JWM technique was also dubbed
{\it almost-balanced} weak values (ABWV) amplification \cite{ABWV16,ABWV17a,ABWV17b}.
In short, the ABWV technique
utilizes two balanced weak values, rather than using
the extremely unbalanced weak value as in the WVA scheme.

Existing theoretical analysis and experimental explorations
revealed the main advantages of the ABWV technique as follows.
{\it (i)}
By subtracting the PSA and PSR readouts,
a {\it WVA-like response} can be obtained in the {\it difference signal}.
In experimental demonstrations,
it was shown that the effect of signal amplification using the ABWV technique
is more prominent than using the WVA technique \cite{ABWV17a,ABWV17b,Zeng16}.
{\it (ii)}
Viewing that in the ABWV scheme
all the readout data are collected without discarding,
it was concluded that this scheme collects all of the
Fisher information of the estimated parameter \cite{Bru13,ABWV16}.
{\it (iii)}
Owing to subtracting, the ABWV technique permits
the removal of systematic error, background noise, and fluctuations in alignments
of the experimental setup \cite{ABWV17a,ABWV17b,Zeng16}.
In Ref.\ \cite{ABWV17b}, it was shown that the ABWV technique
offers on average a twice better signal-to-noise ratio (SNR)
than WVA for measurements of linear velocities.
In Ref.\ \cite{Zeng19},
it was estimated and demonstrated that the JWM has
a sensitivity two orders of magnitude higher than the WVA,
under some technical imperfections (e.g. misalignment errors).
{\it (iv)}
In the ABWV scheme,
prior information about the input state of the pointer is not required,
since the sum of the PSA and PSR outputs can be used as well \cite{ABWV16,ABWV17a,ABWV17b}.

Viewing that the advantages summarized above were largely
based on considerations of some external factors
(e.g. limitations of specific experimental setups),
in this work we present an analysis
for the intrinsic quality of the JWM technique,
in parallel with the analysis for the WVA scheme \cite{Ked12,Jor14,Li20}.
We notice that this type of analysis for the JWM is still lacking in literature.
In particular, the key feature of the JWM technique is using the difference signal,
i.e., the difference of the PSA and PSR probability distribution functions (PDFs).
However, we cannot use this difference PDF to calculate
variance and Fisher information (FI), since it is not positive-definite.
To overcome this difficulty, we introduce the corresponding
difference of combined stochastic variables (DCSVs),
which makes the calculation of the signal amplification and its variance well defined.
We find that, in general, i.e., with the number of the PSA results ($N_1$)
unequal to that of the PSR results ($N_2$),
the JWM scheme cannot reach the metrological precision
indicated by the total FI,
$F^{(N)}_{\rm tot}=N_1F_1 + N_2F_2$,
discussed in Refs.\ \cite{Bru13,ABWV16,ABWV17a,ABWV17b}
based on the fact that all the readouts are collected without discarding.
Moreover, we analyze the effect of technical noise.
Since the JWM technique is using a signal from the combination of two weak-values,
the variance is the statistical sum of the variances of the PSA and PSR readouts.
Therefore, the technical noise cannot be removed by the subtracting procedure
in the JWM scheme. However, by performing the imaginary-WV measurement,
the technical noise can be avoided or can be even utilized \cite{Ked12,Jor14,Li20}.

The paper is organized as follows.
In Sec.\ II we outline the JWM scheme and point out
that the difference probability function
employed by the JWM technique cannot be used to calculate variance and FI.
In Sec.\ III, in terms of DCSV,
we reformulate the problem and make all calculations well defined.
In Sec.\ IV
we analyze the effect of technical noise.
Finally, we summarize the work with brief discussions in Sec.\ V.

\section{Joint-Weak-Measurement Scheme}

Let us consider a two-state system coupled to a meter for JWM,
as schematically shown in Fig.\ 1.
The two-state system can correspond to the electron spin in the Stern-Gerlach setup,
the photon polarization or which-path degree of freedom
in quantum optics experiments, and many other possible realizations.
In the Stern-Gerlach setup, the electron's trajectory is deflected
when it passing through the inhomogeneous magnetic field.
For quantum measurement in this setup \cite{AAV88,AV90},
the interaction between the system (spin) and meter
can be described by $H'=\kappa P A$,
with $P$ the momentum operator and $A=\sigma_z$ the Pauli operator for the spin.
In general, we assume that the spin of the electron
is initially prepared in a quantum superposition
\bea
|i\ra= c_1 |1\ra + c_2 |2\ra \,,
\eea
with $|1\ra$ and $|2\ra$ denoting the spin-up and spin-down states.
The electron's transverse spatial wavefunction
is assumed as a Gaussian
\bea
\Phi(x)=\frac{1}{(2\pi \sigma^2)^{1/4}}
\exp\left[-\frac{x^2}{4\sigma^2}\right],
\eea
with $\sigma$ the width of the wavepacket.
After passing through the region of the inhomogeneous magnetic field,
the entire state of the electron becomes entangled and is given by
\bea\label{WF-T}
|\Psi_T\ra =
c_1 |1\ra |\Phi_1\ra + c_2  |2\ra |\Phi_2\ra \,,
\eea
where the meter's wavefunctions read as
\bea\label{Phijx}
\Phi_j(x)=\frac{1}{(2\pi \sigma^2)^{1/4}}
      \exp\left[-\frac{(x-\bar{x}_j)^2}{4\sigma^2} \right] \,.
\eea
with $\bar{x}_{1,2}=\pm d$ the Gaussian centers
shifted by the coupling interaction $e^{-id P A}$,
where $d=\int_0^{\tau} dt\, \kappa=\kappa \tau$
and $\tau$ is the interacting time.
The parameter $d$ is what we are interested in
and want to estimate
through measuring the spatial wavefunction.

Let us first consider the conventional scheme of measurement,
which does not involve measurement (post-selection)
for the spin state.
In this case,
ignoring the spin state corresponds to tracing the spin degree of freedom,
leaving thus the meter state given by
\bea
\rho_{\rm m} = {\rm Tr}_{\rm s}(|\Psi_T\ra \la\Psi_T|)
= |c_1|^2 |\Phi_1\ra \la\Phi_1| + |c_2|^2 |\Phi_2\ra \la\Phi_2| \,.
\eea
Based on this state, we have $\bar{x} = (|c_1|^2-|c_2|^2)d$.
It becomes clear that, for the conventional measurement (CM) without post-selection,
the optimal choice is to prepare the spin in one of the basis states,
e.g., in the spin-up state $|1\ra$.
This results in the largest signal $\bar{x}=d$.
Meanwhile, the variance (uncertainty) is $\sigma$.
One can thus define the so-called signal-to-noise ratio (SNR),
$R^{(1)}_{\rm CM} = d/\sigma$,
to characterize the quality of measurement.
If using $N$ particles for the measurement (to estimate the parameter $d$),
the estimate precision is characterized by
\bea\label{SNR-CM}
R^{(N)}_{\rm CM} = \frac{d}{\sigma/\sqrt{N}}   \,.
\eea
This SNR will be used as a standard
for comparison with other measurement schemes,
i.e., the WVA and JWM schemes to be discussed in this work.

\begin{figure}
\includegraphics[scale=0.5]{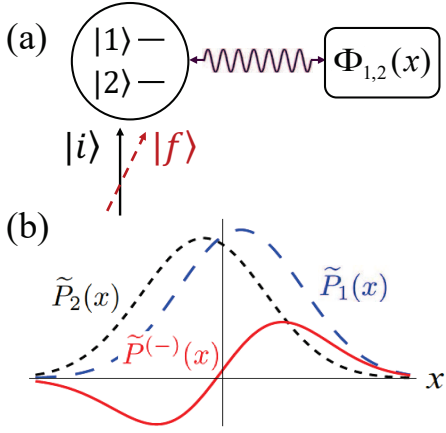}
\caption{(a)
Schematic plot of a two-state system coupled to a meter in joint-weak-measurement (JWM) setup.
The two-state system can correspond to the electron spin in the Stern-Gerlach setup,
the photon polarization or which-path degree of freedom
in quantum optics experiments, and many other possible realizations.
The meter's wavefunction can be such as
the transversal spatial wavefunction of an electron or a photon,
which is deformed (shifted) from $\Phi(x)$ to $\Phi_{1,2}(x)$
by the system states $|1\ra$ and $|2\ra$, owing to measurement coupling interaction.
In the WVA or JWM scheme,
the system is initially prepared (preselected) in state $|i\ra$
and post-selected using $|f\ra$,
while the meter's wavefunction
is projectively measured to output the classical result of $x$.
(b)
The probability distribution functions (un-normalized)
$\widetilde{P}_1(x)$ and $\widetilde{P}_2(x)$
of the PSA and PSR results of $x$.
Their difference, $\widetilde{P}^{(-)}(x)$,
is the probability function
employed in the JWM scheme to estimate the coupling interaction parameter,
which can reveal the WVA-like signal with huge amplification.
However, as schematically shown in the plot,
$\widetilde{P}^{(-)}(x)$ is not {\it positive-definite},
which cannot be used to calculate the distribution variance and Fisher information.
This fundamental difficulty will be solved in this work
by a technique of alternatively introducing
difference-combined stochastic variables (DCSV),
rather than using the ill-defined
probability function $\widetilde{P}^{(-)}(x)$.      }
\end{figure}

Next, let us consider the WVA scheme, in the the AAV limit.
It can be proved that after post-selection with $|f\ra$ for the spin state,
the meter's wavefunction is approximately given by \cite{AAV88,AV90}
\bea\label{WV-Px}
\Phi_f(x)=\frac{1}{(2\pi \sigma^2)^{1/4}}
\exp\left[-\frac{(x- A^f_w\, d)^2}{4\sigma^2} \right]  \,,
\eea   
where the AAV WV reads as
\bea
A^f_w=\frac{\la f|A|i\ra}{\la f|i\ra} \,.
\eea
From the rigorous result of \Eq{wv-1},
we know that the validity condition of \Eq{WV-Px}
is $(d/2\sigma)|A^f_w|<<1$.
For the task of parameter $d$ estimation,
we need the result of average of $x$, in the AAV limit which is simply the center value
of the wavepacket $|\Phi_f(x)|^2$. We see then that, in the WVA scheme,
the signal is enhanced as $\bar{x}=({\rm Re}A^f_w)\, d$,
noting that the AAV WV $A^f_w$ can be very large
(strongly violating the bounds of the eigenvalues of $A$).
Therefore, the SNR of the WVA measurement is characterized by
$R^{(1)}_{\rm WVA} =({\rm Re}A^f_w)d /\sigma $.
When considering $N$ particles used for the measurement
but only $N_1$ particles survived in the post-selection,
the SNR is given by
\bea\label{SNR-WVA}
R^{(N)}_{\rm WVA} = \frac{({\rm Re}A_w)d}{\sigma/\sqrt{N_1} }   \,.
\eea
From this result, despite that the signal is enhanced from $d$ to $({\rm Re}A^f_w)d$,
the SNR of the WVA scheme cannot be much improved as naively expected,
since the uncertainty of estimation grows as well at the same time,
owing to the small probability of successful post-selection,
$p_f\simeq |\la f|i\ra|^2\simeq N_1/N$ under the AAV limit.
It can be proved \cite{Jor14}
that the SNR of WVA scheme, $R^{(N)}_{\rm WVA}$,
can at most reach $R^{(N)}_{\rm CM}$ of the conventional measurement,
if using the same number of particles.
It is just this trade-off consideration that has caused debates in literature
\cite{Nish12,Ked12,Jor14,Li20,Tana13,FC14a,Kne14,ZLJ15,Aha15,Li16,FC14b},
despite that the WVA scheme does have some practical advantages
as demonstrated by experiments \cite{Lun17,Zen19,ZLJ20}.

Now, let us briefly reformulate the JWM scheme as follows.
Beyond the AAV limit, the meter's wavefunctions
associated with success and failure of the post-selection can be expressed as
\bea
\Phi_f(x)&=& \la f;x|\Psi_T\ra
=  c_{1f} \Phi_1(x) + c_{2f} \Phi_2(x)    \,,  \nl
\Phi_{\bar{f}}(x)&=& \la \bar{f};x|\Psi_T\ra
= c_{1\bar{f}}\Phi_1(x) +  c_{2\bar{f}}\Phi_2(x)  \,.
\eea
Here, $|\bar{f}\ra$ is orthogonal to the state $|f\ra$
and the superposition coefficients are updated as:
$c_{1f}=\la f|1\ra c_1$ and $c_{2f}=\la f|2\ra c_2$,
conditioned on success of the post-selection with state $|f\ra$;
$c_{1\bar{f}}=\la \bar{f}|1\ra c_1$
and $c_{2\bar{f}}=\la \bar{f}|2\ra c_2$,
corresponding to failure of the post-selection.
Respectively, the distribution probabilities
of the measurement results are given by
\bea\label{P-fun-1}
\widetilde{P}_1(x)&=&|\Phi_f(x)|^2 \equiv p_f P_1(x)  \,, \nl
\widetilde{P}_2(x)&=&|\Phi_{\bar{f}}(x)|^2 \equiv p_{\bar{f}} P_2(x) \,.
\eea
Here we introduced the normalized probability functions
$P_1(x)$ and $P_2(x)$,
and the probabilities of post-selection success and failure
$p_f$ and $p_{\bar{f}}$,
which are also the normalization factors, i.e.,
$p_f=\int dx\, \widetilde{P}_1(x)$
and $p_{\bar{f}}=\int dx \, \widetilde{P}_2(x)$.

The JWM scheme suggests using
the difference of the probability distribution functions (PDFs),
i.e., $\widetilde{P}^{(-)}(x)= \widetilde{P}_1(x)-\widetilde{P}_2(x)$,
as a signal function from which the parameter is extracted.
To gain an intuitive insight,
and in parallel to the experiments \cite{ABWV16,ABWV17a,ABWV17b,Zeng16},
let us consider the specific pre- and post-selection states
$|i\rangle = (|1\rangle +|2\rangle)/\sqrt{2}$
and $|f\rangle = \cos\frac{\theta}{2}|1\rangle +\sin\frac{\theta}{2}|2\rangle$.
The state orthogonal to the post-selection is
$|\bar{f}\rangle =\sin\frac{\theta}{2}|1\rangle -\cos\frac{\theta}{2}|2\rangle$.
Then, we obtain $\widetilde{P}_1(x)=|\la f;x|\Psi_T\ra|^2$
and $\widetilde{P}_2(x)=|\la \bar{f};x|\Psi_T\ra|^2$,
where $|\Psi_T\ra =(|1\ra |\Phi_1\ra +  |2\ra |\Phi_2\ra)/\sqrt{2}$,
based on \Eq{WF-T}.
The difference probability function is further obtained as \cite{ABWV16}
\bea\label{P-profile}
  \widetilde{P}^{(-)}(x)
   &=& \frac{1}{2}\left(\cos\theta e^{\frac{dx}{2\sigma^2}}-\cos\theta e^{-\frac{dx}{2\sigma^2}}
  +2\sin\theta\right)P(x) \nl
   &\simeq& \sin\theta\,\left(1+\frac{d\cot\theta x}{\sigma^2}\right)P(x) \nl
   &\simeq& \sin\theta \, P(x-d\cot\theta)  \,.
\eea
In deriving this elegant result,
the conditions $d x/\sigma^2<<1$ and $ d\cot\theta x/\sigma^2 <<1$ are used,
which ensure the approximations of
$e^{\pm dx/2\sigma^2}\simeq 1\pm dx/2\sigma^2$
and $1+d\cot\theta x/\sigma^2 \simeq e^{d\cot\theta x/\sigma^2}$.
Here we see that this difference PDF
has a shifted peak similar to WVA \cite{ABWV16},
with the peak center proportional to the parameter $d$ under estimation.
For small $\theta$, $\cot\theta\simeq \frac{1}{\theta}$,
this shift is anomalously amplified.
In this case, one can check that the weak values
$A^{f}_w= \la f|A|i\ra / \la f|i\ra$ and
$A^{\bar{f}}_w= \la {\bar{f}}|A|i\ra / \la {\bar{f}}|i\ra$ are almost equal.
For this reason, the JWM scheme was also referred to as
ABWV technique \cite{ABWV16,ABWV17a,ABWV17b}.

To connect with experiment, let us consider using $N$ particles.
As schematically shown in Fig.\ 1, the PSA and PSR PDFs correspond to
\bea
\widetilde{P}_1(x) = \frac{n_1(x)}{N}
~~{\rm and}~~ \widetilde{P}_2(x) = \frac{n_2(x)}{N} \,,
\eea
where $n_1(x)$ and $n_2(x)$ are
the numbers of the PSA and PSR particles at point $x$.
Then, one can define the {\it difference signal} as
\bea\label{diff-P-1}
P^{(-)}(x)=\frac{n_1(x)-n_2(x)}{N_1-N_2}   \,.
\eea
This is the normalized version of $\widetilde{P}^{(-)}(x)$,
with $N_1$ and $N_2$ the total PSA and PSR particle numbers.
Using $P^{(-)}(x)$, one can estimate the parameter $d$
from the average $\bar{x} = \int dx\, x\, P^{(-)}(x)$, which reads
\bea\label{xbar-1}
\bar{x}
= \la x\ra_{f} \left(\frac{\delta_1}{\delta_1-\delta_2}\right)
- \la x\ra_{\bar{f}} \left(\frac{\delta_2}{\delta_1-\delta_2}\right)  \,,
\eea
where $\delta_1=N_1/N$ and $\delta_2=N_2/N$
correspond to the post-selection success and failure probabilities
$p_f$ and $p_{\bar{f}}$.
In experiment, the conditional averages $\la x\ra_{f}$ and $\la x\ra_{\bar{f}}$
can be determined using the distribution functions
$P_1(x)=n_1(x)/N_1$ and $P_2(x)=n_2(x)/N_2$;
while in theory, they are computed using the normalized probability functions
$P_1(x)$ and $P_2(x)$ given by \Eq{P-fun-1}.
Then, making contact between the experimental and theoretical results of $\bar{x}$,
one can extract (estimate) the value of the parameter $d$.
The theoretical result becomes quite simple in the AAV limit,
$\la x\ra_{f}={\rm Re}A^f_w d$ and $\la x\ra_{\bar{f}}={\rm Re}A^{\bar{f}}_w d$.
Substituting them into \Eq{xbar-1},
we know how to extract the parameter $d$
from the experimental result of $\bar{x}$.
Actually, in practice, even simpler method is possible \cite{ABWV16,ABWV17a,ABWV17b}:
using the collected data to fit the difference PDF (wavepacket profile)
such as \Eq{P-profile}, one can extract the relevant parameters.

For arbitrary strength of measurement (beyond the AAV limit),
using $P_1(x)$, one can straightforwardly obtain
the WVA amplification factor as \cite{Li20,Wu11,Tana11,Naka11}
\begin{equation}\label{wv-1}
 \frac{ \langle x\rangle_f }{d}
  = \frac  {{\rm Re}A^f_w}{1+ {\cal G}\, (|A^f_{w}|^{2}-1)}
  \equiv \frac{{\rm Re}A^f_w}{{\cal M}_1}   \,.
\end{equation}
Here we introduced ${\cal G}=(1-e^{-2g})/2$ and $g=(d/2\sigma)^2$,
which is a suitable parameter to characterize the measurement strength.
We also defined the modification factor ${\cal M}_1$,
which clearly reflects the modification effect to the AAV result.
Using $P_2(x)$, similar result can be obtained for $ \la x\ra_{\bar{f}}\,/\,d$,
by only replacing $A^f_{w}$ with $A^{\bar{f}}_{w}$
and denoting the modification factor as ${\cal M}_2$.

The WVA scheme utilizes only the signal $\la x\ra_{f}$
in the first term of \Eq{xbar-1}.
Through proper choice of the post-selection state $|f\ra$
(making it nearly orthogonal to $|i\ra$),
one can obtain an anomalous AAV WV,
which corresponds to an anomalously large $\la x\ra_{f}$.
This is the basic principle of WVA.
From \Eq{xbar-1}, the amplification principle of the JWM scheme is quite different:
the anomalous amplification is owing to $\delta_1\simeq\delta_2$.
We know that the anomalous WV is deeply rooted in
a nature of quantum interference \cite{Li16}.
For classical systems, it is impossible to realize such type of amplification.
However, from \Eq{xbar-1}, the amplification principle in the JWM scheme
is seemingly applicable to measurements and statistics in classical systems.

In Fig.\ 2, we compare the amplification effects of the JWM and WVA techniques.
For arbitrary strength of measurement,
the JWM scheme can realize anomalously large amplification
in the ABWV regime with $N_1\simeq N_2$.
However, for WVA, only at the AAV limit, it is possible to make
the amplification factor anomalously large.
With increase of the measurement strength,
the amplification effect will be weakened \cite{Li20,Wu11,Tana11,Naka11}.

\begin{figure}
\includegraphics[scale=0.35]{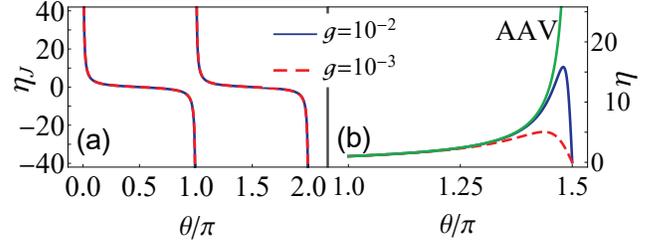}
\caption{ Amplification factors of JWM in (a) and WVA in (b),
defined by $\bar{x}/d$, by noting that $d$ is
the center's shift of the meter wavefunction in conventional measurement.
For JWM,
the amplification effect is not sensitive to the measurement strength.
For finite $g$ beyond the AAV limit,
the singular behavior can exist in the ABWV regime,
i.e., $A^f_w\simeq A^{\bar{f}}_w$ when $\theta\to \pi$.
In contrast, for WVA,
the singular behavior of $\eta={\rm Re}A^f_w$ at the AAV limit
(plotted by the green line in (b))
when $\la f|i\ra\to 0$ and the anomalous amplification will be
suppressed with increase of the measurement strength $g$
\cite{Li20,Wu11,Tana11,Naka11}.  }
\end{figure}

The amplification in the JWM is realized through
the mean value of $x$ governed by $P^{(-)}(x)$.
However, using $P^{(-)}(x)$, we cannot calculate the distribution variance,
since it is not {\it positive-definite}, as illustrated in Fig.\ 1(b).
Also, a non-positive-definite probability function
will render the Cr\'amer-Rao bound (CRB) inequality meaningless:
it does not allow us to infer the estimate precision
from the ``Fisher information" computed using it.
Actually, in this case, the ``Fisher information" itself is problematic.
To be more general, let us consider
an arbitrary parameter-$\lambda$-dependent probability function $P(x;\lambda)$,
we have $\int dx P(x)(x-\bar{x}) = 0$.
Here, for brevity, we have denoted $P(x;\lambda)$ by $P(x)$
and $\bar{x}$ is the average of $x$ determined by $P(x;\lambda)$.
Straightforwardly,
making derivative with respect to the parameter $\lambda$, we have
\bea
\int dx \left( \frac{\partial P(x)}{\partial \lambda} \right) (x-\bar{x})
= \frac{\partial \bar{x}}{\partial \lambda}  \,.
\eea
Further, applying the Cauchy-Schwarz inequality yields
\bea\label{CRB-1}
\left(\frac{\partial \bar{x}}{\partial \lambda}\right)^2
\le {\cal F}_{\lambda}\, \int dx\, |P(x)| (x-\bar{x})^2   \,,
\eea
where
${\cal F}_{\lambda} = \int dx \,|P(x)|
\left( \frac{1}{P(x)}   \frac{\partial P(x)}{\partial \lambda} \right)^2$.
For positive-definite probability function, every thing is perfect.
That is, ${\cal F}_{\lambda}$ is the Fisher information about the parameter $\lambda$,
which satisfies the so-called CRB inequality
$\left(\frac{\partial \bar{x}}{\partial \lambda}\right)^2 / {\rm D}[x]
\leq {\cal F}_{\lambda}$,
where the variance ${\rm D}[x]=\int dx P(x)(x-\bar{x})^2$.
In the special case of $\bar{x}=\lambda$,
i.e., $x$ being the unbiased estimator of $\lambda$,
it reduces to the standard form of the CRB inequality,
$\frac{1}{{\rm D}[x]}\leq  {\cal F}_{\lambda}$.

For a non-positive-definite probability function, such as the difference PDF $P^{(-)}(x)$,
it seems that one can still compute the ``Fisher information" using the above formula.
However, in the inequality \Eq{CRB-1}, the ``statistical variance" of $x$ becomes meaningless.
We notice that the Fisher information encoded in the JWM was suggested as \cite{ABWV16}
\bea\label{F-tot}
F^{(N)}_{\rm tot} = N_1 F_1 + N_2 F_2  \,,
\eea
where the meaning of the particle numbers $N_1$ and $N_2$
is the same as above,
while $F_1$ and $F_2$ are the Fisher information
associated with the PSA and PSR PDFs, i.e., $P_1(x)$ and $P_2(x)$, respectively.
As we will see in the following section (Fig.\ 4), using this total FI
will overestimate the metrological precision of the JWM,
in comparison with the result we calculate directly
through a standard statistical method based on introducing stochastic variables.

\section{Reformulation in terms of Stochastic Variables}

Instead of using the difference probability function $P^{(-)}(x)$,
let us introduce the statistical description in terms of stochastic variables.
From the viewpoint of probability theory, each result ``$x_j$" of measurement
is a specific realization of the stochastic variable ``$\hat{x}_j$",
which is governed by the property of the underlying statistical ensemble.
Let us consider grouping the stochastic variables as follows
\bea
\hat{Y}_1 &=& \frac{1}{N_1} \sum^{N_1}_{j=1} \hat{x}^{(f)}_j  \,,  \nl
\hat{Y}_2 &=& \frac{1}{N_2} \sum^{N_2}_{k=1} \hat{x}^{(\bar{f})}_k  \,.
\eea
This corresponds to the experiment using $N$ particles,
in which there are $N_1$ particles accepted by the post-selection,
and $N_2$ particles rejected.
In the first group,
each stochastic variable obeys the statistics
governed by $P_1(x)$ defined in \Eq{P-fun-1},
while in the second group
each stochastic variable obeys the statistics
governed by $P_2(x)$.
Then, the difference signal (the average $\bar{x}$) exploited in the JWM scheme
corresponds to {\it a single realization}
of the following difference-combined stochastic variable (DCSV)
\bea\label{SV-1}
\hat{x}&=&\left(\frac{N_1}{N_1-N_2}\right) \hat{Y}_1
- \left(\frac{N_2}{N_1-N_2}\right) \hat{Y}_2 \nl
&\equiv& ~ \beta_1 \hat{Y}_1 -\beta_2 \hat{Y}_2   \,.
\eea
The ensemble average of $\hat{x}$ reads as
\bea\label{SV-2}
{\rm E}[\hat{x}] = \beta_1 \la x\ra_f - \beta_2 \la x\ra_{\bar{f}} \,,
\eea
which is the same as the $\bar{x}$ given by \Eq{xbar-1}
using the difference probability function $P^{(-)}(x)$,
by noting that the ratio factors
$\delta_1/(\delta_1-\delta_2)$ and $\delta_2/(\delta_1-\delta_2)$ in \Eq{xbar-1}
are just the factors $\beta_1$ and $\beta_2$ introduced above in \Eq{SV-1}.
However, the variance of $\hat{x}$
\bea\label{SV-3}
{\rm D}[\hat{x}] &=& \beta^2_1\, {\rm D}[\hat{Y}_1] + \beta^2_2\, {\rm D}[\hat{Y}_2]   \nl
&=& \beta^2_1 \left(\frac{\sigma^2_1}{N_1}\right)
    + \beta^2_2 \left(\frac{\sigma^2_2}{N_2}\right)   \,,
\eea
is now well-defined and positive-definite,
which properly characterizes the estimate precision of the JWM.
Here $\sigma^2_1$ and $\sigma^2_2$ are
the variances of the single stochastic variables
in the sub-ensembles defined by $P_1(x)$ and $P_2(x)$, respectively,
which are introduced in \Eq{P-fun-1}.
Under the AAV limit, both $P_1(x)$ and $P_2(x)$ are still Gaussian functions,
with shifted centers of ${\rm Re}A^f_w d$ and ${\rm Re}A^{\bar{f}}_w d$
but keeping the widths $\sigma_1=\sigma_2=\sigma$ unchanged, as shown by \Eq{WV-Px}.
For finite strength of measurement, however, they are no longer
Gaussian in general and may have different widths.
Using $P_1(x)$ and $P_2(x)$ to calculate the variances, we obtain \cite{Li20}
\bea\label{var-1}
\sigma_1^2 &=& \langle x^2 \rangle_f - (\langle x \rangle_f)^2  \nl
&=&  \sigma^2 + d^2 \eta \left(  \frac{|A^f_w|^2+1}{2{\rm Re}A^f_w} -\eta \right)  \,.
\eea
Here we introduced the amplification factor $\eta = |\la x\ra_f|/d$.
Similar result of $\sigma_2^2$
can be obtained by replacing $A^f_w$ with $A^{\bar{f}}_w$,
and $\eta$ by $\bar{\eta}=|\la x\ra_{\bar{f}}|/d$.

From the above results, we know that,
with the signal being amplified, its fluctuation increases.
Thus, a reasonable characterization to
the metrological quality of the JWM scheme is still the SNR
\bea\label{SNR-JWM}
R^{(N)}_{\rm JWM} = \frac{\beta_1 \la x\ra_f - \beta_2 \la x\ra_{\bar{f}}}
{\sqrt{ \beta^2_1 \left(\frac{\sigma^2_1}{N_1}\right)
    + \beta^2_2 \left(\frac{\sigma^2_2}{N_2}\right)  }}  \,.
\eea
In Fig.\ 3, we plot this SNR
as a function of the post-selection angle $\theta$ and the measurement strength $g$,
and compare it with the SNR of the WVA.
For both techniques, the maximum of the SNR are bounded by
$\frac{d}{\sigma/\sqrt{N}}$, i.e., the SNR of conventional scheme.
For the JWM technique,
the maximum SNR can always be achieved
in the ABWV regime with $N_1\simeq N_2$.
Similar as the amplification factor,
this behavior of SNR is not sensitive to the measurement strength.
In contrast, for WVA,
we find that for small $g$, large range of post-selection
can reach the maximum SNR.
However, with the increase of $g$,
this range is narrowed and gradually disappears.

\begin{figure}
\includegraphics[scale=0.3]{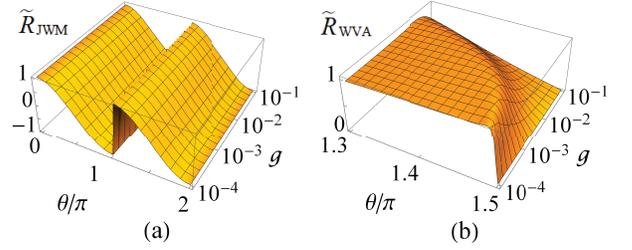}
\caption{ Signal-to-noise ratio (SNR) of JWM in (a) and WVA in (b).
Both results are scaled by the SNR of conventional measurement,
$R^{(N)}_{\rm CM}=\sqrt{N}d/\sigma$,
which is found to set a upper bound for the SNR of both JWM and WVA.
For JWM, the SNR of anomalous amplification is not sensitive to $g$.
In contrast, for WVA,
the SNR associated with the anomalous amplification measurement
is reduced with increase of the measurement strength $g$.  }
\end{figure}

As explained already, we cannot calculate the Fisher information
using the difference probability function $P^{(-)}(x)$.
However, based on the DCSV $\hat{x}$, which gives the well-defined
expectation value (the amplified signal) $\widetilde{d}={\rm E}[\hat{x}]$
and the $\widetilde{d}$-estimation uncertainty
$\delta(\widetilde{d})=({\rm D}[\hat{x}])^{1/2}$,
we can define the stochastic variable $\hat{y}=\hat{x}/\widetilde{\eta}$,
where $\widetilde{\eta}=\widetilde{d}/d$
is the amplification factor of the JWM scheme.
Straightforwardly, we have ${\rm E}[\hat{y}]=d$
and ${\rm D}[\hat{y}]={\rm D}[\hat{x}]/\widetilde{\eta}^2$,
which are, respectively, the expectation value and variance of the parameter-$d$-estimation.
Then, we propose to define the following effective FI for the JWM
\bea\label{eff-FI}
F^{(N)}_{\rm JWM}= \frac{({\rm E}[\hat{x}]/d)^2}{{\rm D}[\hat{x}]} \,,
\eea
to be used to compare with the FI of the WVA and in particular,
with the total FI given by \Eq{F-tot}.
In Fig.\ 4, we find that, for very small $g$ (in the AAV limit),
the WVA technique can reach the FI of conventional measurement
in the regime of near orthogonal post-selection (i.e., $\la f|i\ra\simeq 0$).
The novel point in this context is that the post-selected small portion
of measurement data contains almost the full information,
which leads thus to a practical advantage
in the presence of power saturation of the detectors.
With the increase of $g$ (away from the AAV limit), we find that
the WVA technique cannot reach the FI of conventional measurement.
For the JWM scheme, in the ABWV regime with $N_1\simeq N_2$,
the effective FI can reach the FI of conventional measurement,
regardless of the measurement strength $g$.
The reason for this difference is as follows.
The WVA scheme employs the anomalously large
conditional average $\la x\ra_f$ as the amplified signal,
which is rooted in a nature of quantum interference \cite{Li16}.
With the increase of $g$, this amplification effect will be reduced.
On the contrary, the JWM scheme is free from quantum interference effect.

An important observation here is that,
except for the special post-selection with $N_1\simeq N_2$,
the effective FI of the JWM scheme
drastically deviates from
the total FI $F^{(N)}_{\rm tot}$ given by \Eq{F-tot}.
This implies that, if using $F^{(N)}_{\rm tot}$,
the metrology precision of the JWM
will be overestimated for the case of $N_1\neq N_2$.
That is, the estimate uncertainty determined by $1/F^{(N)}_{\rm tot}$
is smaller than ${\rm D}[\hat{x}]/\widetilde{\eta}^2$
from the direct calculation based on the DCSV technique.

We find that the optimal results
are achieved in the ABWV regime with $N_1\simeq N_2$.
If the post-selection is not set in this regime,
e.g., with $N_1\simeq \beta N_2$,
one may consider an asymmetric combination of the PSA and PSR data.
In terms of probability distribution function, we have
\bea\label{diff-P-2}
P_{\beta}^{(-)}(x)=\frac{n_1(x)-\beta\, n_2(x)}{N_1-\beta N_2}   \,,
\eea
Similar to the symmetric combination scheme discussed above,
we can construct the DCSV as
\bea
\hat{x}_{\beta}&=&\left(\frac{N_1}{N_1-\beta N_2}\right) \hat{Y}_1
- \left(\frac{\beta N_2}{N_1-\beta N_2}\right) \hat{Y}_2 \nl
&\equiv& ~ \tilde{\beta}_1 \hat{Y}_1 -\tilde{\beta}_2 \hat{Y}_2   \,.
\eea
Based on this DCSV, we can carry out the SNR characterization as well.
One can easily check that,
for both the amplification factor and the SNR,
the same optimal results can be achieved
by post-selection satisfying $N_1\simeq \beta N_2$,
instead of the ABWV with $N_1\simeq N_2$.
This conclusion is valid also for the results
in the presence of technical noise,
which is to be addressed in the following.

\begin{figure}
\includegraphics[scale=0.45]{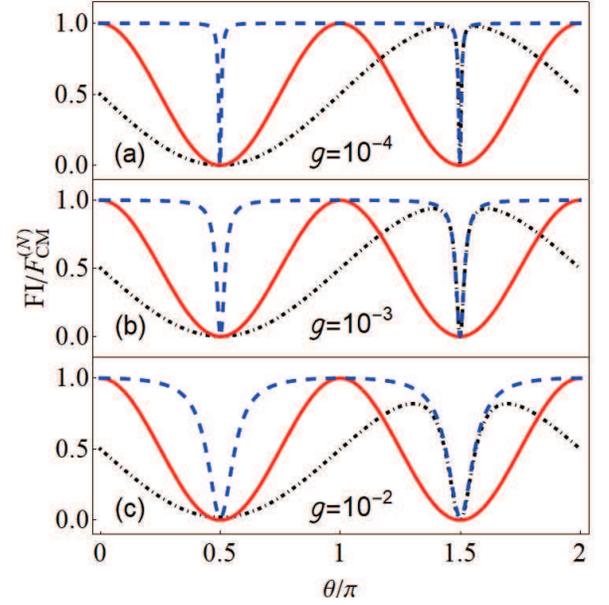}
\caption{Effective Fisher information (FI) of JWM (solid red)
and the FI of WVA (dash-dotted black),
compared with the total FI $F^{(N)}_{\rm tot}=N_1F_1 + N_2F_2$ (dashed blue).
All results are scaled with the FI of conventional measurement,
$F^{(N)}_{\rm CM}=N /\sigma^2$.
For both the WVA at the AAV limit
and the JWM in the ABWV regime,
the FI can reach the result of conventional measurement.
For most range of post-selection, $F^{(N)}_{\rm tot}$
coincides also with $F^{(N)}_{\rm CM}$.
However, for JWM, except for the ABWV case with $N_1\simeq N_2$,
the effective FI drastically deviates from
$F^{(N)}_{\rm tot}$ which was regarded
as the FI collected by the JWM technique \cite{Bru13,ABWV16}.   }
\end{figure}

\section{Effect of Technical Noise}

In practice, there may exist technical issues to cause extra noises,
which are usually referred to as technical noises.
In general, technical noises will increase the uncertainty of estimation,
being thus harmful to precision metrology.
For the real-WV measurement as shown by \Eq{WV-Px} at the AAV limit,
it can be proved that the technical noise does not cause
any shift of the signal (i.e., the average of $\la x\ra_f$),
but does increase the variance of $x$.
Therefore, the effect of the noise is similar as
in the conventional measurement without post-selection.
However, the story is quite different in the imaginary-WV measurement \cite{Li20}.
Given the form of the measurement interaction Hamiltonian $H'=\kappa P A$,
the imaginary-WV measurement can be realized by performing measurement
in the $p$ basis, i.e., in the eigen-basis of the coupling operator $P$,
while the real-WV measurement is performed in the conjugated $x$ basis.
Note also that, it is impossible to perform
the conventional measurement (without post-selection) in the $p$-basis,
since the average of $p$ does not depend on the parameter $d$ under estimation.
Below, we present an investigation
for the effect of technical noise in the JWM scheme.

{\it Measurement in the $x$ basis: $x_0$ noise}.---
Let us first consider the technical noise $x_0$.
The {\it joint probability}
of getting $x$ with the initial state $|i\ra$
and passing the post-selection with $|f\ra$,
and as well with the specific noise $x_0$, is given by
\bea
{\rm Pr}(f;x,x_0)=P_i(x-x_0)\, P_{x-x_0}(f)\, {\rm Pr}(x_0)/{\cal N}_f \,.
\eea
${\cal N}_f$ is a normalization factor.
The first two probability functions read as
$P_i(y)=|c_1|^2|\Phi_1(y)|^2 + |c_2|^2|\Phi_2(y)|^2$
and $P_y(f)= |\la y;f| \Psi_T \ra |^2$.
Here we have denoted $x-x_0$ by $y$ for brevity.
The probability distribution of the noise is assumed to be Gaussian
\bea\label{x0-noise}
{\rm Pr}(x_0)=\frac{1}{\sqrt{2\pi}J} \, e^{-x^2_0/2J^2} \,,
\eea
where $J$ is the width of the noise distribution.
Straightforwardly, one can compute the statistical average through
$ \la \bullet\ra_f = \int dx_0 \int dx \, (\bullet)\, {\rm Pr}(f;x,x_0)$.
We obtain
\bea
&  \la x\ra_f
= \left(\frac{{\rm Re}A^{f}_w}{{\cal M}_1}\right) d  \,,     \nl
&  \la x^2\ra_f
= \sigma^2 + J^2 + \left( \frac{\eta d^2}{2} \right)
\left( \frac{1+|A^{f}_w|^2}{{\rm Re}A^{f}_w} \right)   \,.
\eea
Accordingly, the variance is obtained as usual as
$\sigma^2_{1J}=\la x^2\ra_f-(\la x\ra_f)^2$.
The above results for the PSA data
show that the signal shift
is not affected by the noise, however,
the variance is added by $J^2$, as expected.
Similar results of $\la x\ra_{\bar{f}}$, $\la x^2\ra_{\bar{f}}$
and the variance $\sigma^2_{2J}$ for the PSR data can be obtained,
by replacing $A^{f}_w$, $\eta$ and ${\cal M}_1$
by $A^{\bar{f}}_w$, $\bar{\eta}$ and ${\cal M}_2$, respectively.
Here, we introduced $\bar{\eta}=|\la x\ra_{\bar{f}}|/d$.
Substituting these results into \Eq{SNR-JWM},
the SNR in the presence of the $x_0$ noise can be quantitatively computed.
In this context, we may point out that
the particle numbers $N_1$ and $N_2$
and the ratio parameters $\beta_1$ and $\beta_2$
in \Eq{SNR-JWM} are not affected by the technical noise.
The basic conclusion for the effect of the noise
is the same as for the conventional and WVA schemes:
the technical noise will reduce the estimate precision,
by adding extra uncertainty of $J^2$ as shown above.

{\it Measurement in the $p$ basis: $x_0$ and $p_0$ noises}.---
In order to avoid the influence of the technical noise
and even possibly utilize it, let us consider further
the so-called {\it imaginary} WV measurement \cite{Ked12,Jor14,Li20,Li16}.
This can be realized by performing measurement in the $p$ basis
(see Appendix A for some details),
i.e., in the eigen-basis of the coupling operator $P$
in the measurement interaction Hamiltonian $H'=\kappa P A$.
Therefore, we Fourier-transform the meter's wavefunctions
from the $x$-representation to
\bea
\Phi_{1,2}(p)=\left(\frac{\pi}{2\sigma^2}\right)^{-1/4}\exp[-\sigma^2 p^2  \mp id\,p] \,.
\eea
Let us consider two types of noises.
{\it (i)}
The noise is introduced through $x_0$,
i.e., a random shift of the meter's wavefunction in the $x$ basis.
In this case,
the meter's wavefunctions in the $p$ basis can be reexpressed as
\bea
\Phi_{1,2}(p;x_0)=\left(\frac{\pi}{2\sigma^2}\right)^{-1/4}
\exp[-\sigma^2 p^2  \mp id\,p - ix_0\,p] \,.
\eea
As in the $x$-basis measurement,
the $x_0$ noise satisfies the same Gaussian statistics of \Eq{x0-noise}.
{\it (ii)}
The noise is introduced through a random $p_0$ shift of the $p$ wavepacket.
For a specific $p_0$,
the meter's wavefunctions are shifted from $\Phi_{1,2}(p)$ to
\bea
\Phi_{1,2}(p;p_0)=\left(\frac{\pi}{2\sigma^2}\right)^{-1/4}
\exp[-\sigma^2(p-p_0)^2 \mp i d p]
\eea
As for the $x_0$ noise, the $p_0$ noise is assumed as well a Gaussian
\bea
{\rm Pr}(p_0)=\frac{1}{\sqrt{2\pi}J_p} \, e^{-p^2_0/2J_p^2} \,,
\eea
with $J_p$ the width of the noise distribution.

For both types of noise,
the joint probabilities are given by
\bea
{\rm Pr}(f;p,x_0)&=& P_i(p,x_0)\, P_p(f)\, {\rm Pr}(x_0)/{\cal N}_f \,,  \nl
{\rm Pr}(f;p,p_0)&=& P_i(p,p_0)\, P_p(f)\, {\rm Pr}(p_0)/{\cal N}_f \,.
\eea
${\cal N}_f$ in each result is a normalization factor,
and the first two probability functions read as
$P_i(p)=|c_1|^2|\Phi_1(p)|^2 + |c_2|^2|\Phi_2(p)|^2$
and $P_p(f)= \la f |\widetilde{\rho}(p)|f\ra$.
Here, for brevity, we neglect the noise labels $x_0$ and $p_0$.
In both results, conditioned on the $p$ outcome of measurement,
the spin state is updated as
$\widetilde{\rho}_{12}(p)=\rho_{i12}\, e^{-i\, 2d\, p}$,
while the diagonal elements of the density matrix remain unchanged.
This simply follows the result of
$\widetilde{\rho}(p)=|\widetilde{\psi}(p)\ra\la \widetilde{\psi}(p)|$,
while the $p$-dependent spin state is obtained through
$|\widetilde{\psi}(p)\ra=\la p|\Psi_T \ra
= c_1\Phi_1(p)|1\ra + c_2\Phi_2(p)|2\ra$, based on \Eq{WF-T}.

For the type of $x_0$ noise,
since $|\Phi_1(p;x_0)|^2$ and $|\Phi_2(p;x_0)|^2$
are free from the noise $x_0$,
then $P_i(p,x_0)$ does not depend on $x_0$
and averages of $p$ and $p^2$ are free from $x_0$.
Therefore, an important conclusion is that
the measurement in the $p$ basis
can eliminate the harmful effect of the $x_0$ noise.
Note, however, that this advantage can be possible
only by inserting the ingredient of post-selection measurement.
One can check that, for the $p$ basis measurement
without post-selection (the conventional scheme), the `signal' is zero.
Using the joint probability ${\rm Pr}(f;p,x_0)$, the averages of $p$ and $p^2$
for the PSA data can be obtained as
\bea\label{p-meas-1}
& \la p\ra_f
= \left( \frac{{\rm Im}A^{f}_w}{{\cal M}_1} \right)
  \left(\frac{d}{2\sigma^2}\right)  e^{-d^2/2\sigma^2}  \,, \nl
& \la p^2\ra_f
= \frac{1}{4\sigma^2}  + \left( \frac{|A^{f}_w|^2-1}{{\cal M}_1} \right)
\left(\frac{d^2}{8\sigma^4}\right) e^{-d^2/2\sigma^2}  \,.
\eea
Precisely along the same line,
using the joint probability ${\rm Pr}(\bar{f};p,x_0)$
for the PSR data,
the averages $\la p\ra_{\bar{f}}$ and $\la p^2\ra_{\bar{f}}$ can be obtained.

For the type of $p_0$ noise,
using the joint probability ${\rm Pr}(f;p,p_0)$, the averages of $p$ and $p^2$
for the PSA data can be obtained as
\bea \label{p-meas-2}
& \la p\ra_f
= \left( \frac{{\rm Im}A^{f}_w}{{\cal M}_{1k}} \right)
\,(2d/\widetilde{\sigma}^2_J) \, e^{-2 d^2/\widetilde{\sigma}^2_J}  \,,   \nl
& \la p^2\ra_f
=  1/\widetilde{\sigma}^2_J
 + \left( \frac{|A^{f}_w|^2-1} {{\cal M}_{1k}} \right)
 (2d^2/\widetilde{\sigma}^4_J) e^{-2d^2/\widetilde{\sigma}^2_J}  \,.
\eea
Here we introduced the second modification factor beyond the AAV limit,
${\cal M}_{1k} = 1+ K(|A^{f}_w|^2-1)$,
with $K=(1-e^{-2 d^2/\widetilde{\sigma}^2_J})/2$.
We also introduced an effective width of uncertainty through
\bea\label{width-Jp}
1/\widetilde{\sigma}^2_J =\frac{1}{4\sigma^2} + J^2_p  \,.
\eea
Again, parallel results of
$\la p\ra_{\bar{f}}$ and $\la p^2\ra_{\bar{f}}$
for the PSR data can be obtained.

Knowing $\la p\ra_f$, $\la p\ra_{\bar{f}}$,
and the variances $\sigma^2_{1J}$ and $\sigma^2_{2J}$,
applying \Eq{SNR-JWM} we can compute the SNR in the presence of the $p_0$ noise.
From \Eq{SNR-JWM}, we understand that the effect of the noise
is basically rooted in the ratios
$\la p\ra_f/\sigma_{1J}$ and $\la p\ra_{\bar{f}} /\sigma_{2J}$,
while the more quantitative behaviors are modulated
by the ratios $\beta_1$ and $\beta_2$ of the PSA and PSR readouts.
In this context, we may remark that it is impossible to perform
the conventional measurement (without post-selection) in the $p$-basis,
since the average of $p$ does not depend on the parameter $d$ under estimation.

\begin{figure}
\includegraphics[scale=0.5]{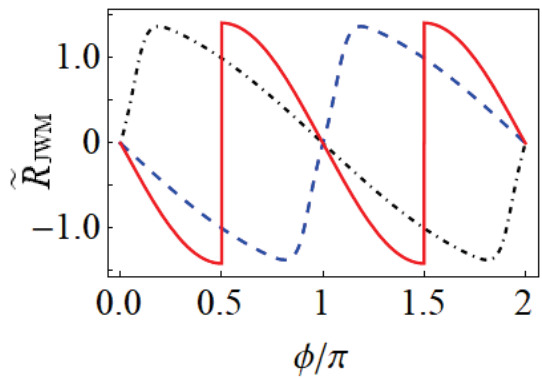}
\caption{ SNR of imaginary weak-value measurement in the presence of technical noise
(the $p_0$-type, see main text for details).
In order to generate imaginary weak values,
with the initial system state $|i\ra=(|1\ra+|2\ra)/\sqrt{2}$,
the post-selected state is designed as
$|f\rangle=(|1\rangle  + e^{-i\phi} |2\rangle) /\sqrt{2}$.
Plotted are the SNR of JWM (solid red),
WVA post-selected by $|f\ra$ (dashed blue)
and WVA post-selected by $|\bar{f}\ra$ (dash-dotted black), respectively.
The state $|\bar{f}\ra$ is orthogonal to $|f\ra$.
All the results are scaled by $R^{(N)}_{\rm CM}=\sqrt{N}d/\sigma$
and arbitrary units with $d=1$ are used.
$J_p=0.1d^{-1}$ and $g=(d/2\sigma)^2=10^{-2}$
are assumed for the noise strength and measurement strength.
The results show that, in the presence of the $p_0$-type noise,
either the JWM or WVA technique
can achieve estimate precision beyond the conventional measurement.
This is the remarkable effect of ``utilizing" technical noise.
Also, the results show that the subtracting procedure of generating
the {\it difference signal} cannot eliminate the technical noise
considered in this work, unlike the situation
considered in Refs.\ \cite{ABWV17a,ABWV17b}.      }
\end{figure}

\begin{figure}
\includegraphics[scale=0.4]{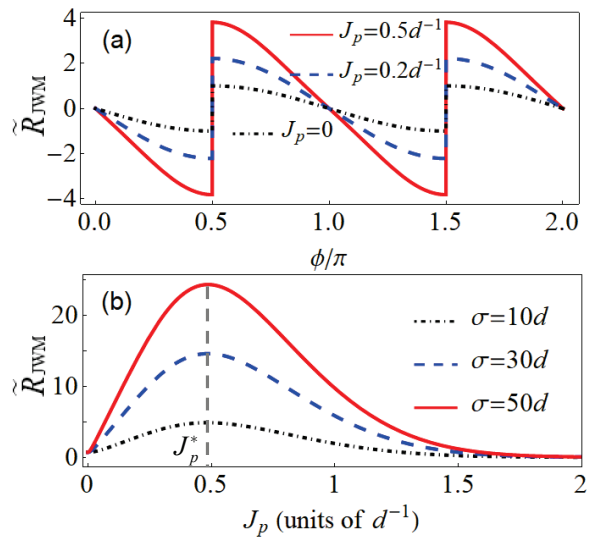}
\caption{ Dependence of the $p_0$ noise strength.
In (a) the complete range of post-selection is shown,
while in (b) the post-selection state with $\phi=3\pi/4$ is considered.
The pre- and post-selected states $|i\ra$ and $|f\ra$ are the same as in Fig.\ 5.
All the results of SNR of the JWM
are scaled by $R^{(N)}_{\rm CM}=\sqrt{N}d/\sigma$.
In (a) the measurement strength $g=(d/2\sigma)^2=10^{-2}$ is assumed
and in (b) the results of more strengths are shown.
From (b), we find that the weaker the measurement strength is,
the larger the noise-assisted enhancement of estimate precision is.
However, this enhancement holds only for noise strength
lower than the critical value $J^*_p$,
as indicated by the vertical dashed line.   }
\end{figure}

In Fig.\ 5 we compare the SNR of JWM with that of WVA,
by varying the post-selection angle $\phi$.
For the JWM,
we find that the ABWV regime with $N_1\simeq N_2$
leads to the peaks at $\phi=\frac{\pi}{2}$ and $\frac{3\pi}{2}$,
i.e., the discontinuous jumps, since we keep the sign of $N_1-N_2$.
We find also that the WVA measurement can reach
similar maximum of SNR as well, at proper post-selection angle,
which does not yet correspond to the post-selection
nearly orthogonal to the initial state.
Actually, we know that the SNR of JWM is related to the SNR of WVA
through \Eq{SNR-JWM}.
This determines their behaviors as shown in Fig.\ 5.
In particular, at $\phi=\frac{\pi}{2}$ and $\frac{3\pi}{2}$,
the ABWV measurement
holds equal SNR for the PSA and PSR results.
In the experiment of Ref.\ \cite{ABWV17b},
under equal conditions for both techniques (WVA and JWM),
it was revealed that
the JWM technique offered a twice better SNR than WVA.
This might originate from the detecting limitations of the WVA,
which make the WVA not working at optimal post-selection.
Finally, we may remark that, as shown in Fig.\ 5 and by \Eq{SNR-JWM},
the subtracting procedure of the JWM technique
cannot remove the technical noise under consideration.

From \Eq{SNR-JWM}, we know that the SNR of JWM
is determined by the SNR of the individual weak-values.
Therefore, similar as the imaginary WVA technique,
the JWM scheme can utilize the technical noise as well, as shown in Fig.\ 6.
In Fig.\ 6(a), we show that the SNR is enhanced
with the increase of the noise (its distribution width $J_p$).
By varying the post-selection angle $\phi$,
the overall behavior is displayed.
Again, we find that the enhancement is most prominent
at $\phi=\frac{\pi}{2}$ and $\frac{3\pi}{2}$,
which correspond to the ABWV regime.
In Fig.\ 6(b), we show the whole $J_p$ dependence, which indicates that
utilizing the noise is possible only for $J_p<J^*_p$.
When $J_p>J^*_p$, the SNR will decrease with increase of the noise strength.
Determination of the critical value $J^*_p$
is referred to the detailed discussion in Ref.\ \cite{Li20}

\section{Summary and Discussion}

We have presented a theoretical analysis
for the metrology quality of the JWM technique
in close comparison with the WVA scheme.
From the aspect of the difference signal (expectation value),
we revealed the {\it classical} origin of anomalous amplification,
which is quite different from the {\it quantum} nature rooted in the WVA scheme.
We pointed out that the difference probability function
cannot be used to calculate the variance,
being thus incapable of characterizing the estimate precision.
Therefore, we reformulate the problem in terms of DCSV,
which makes all calculations well defined.
Our results show that the SNR and FI of the JWM and WVA schemes
are comparable and are bounded by the conventional scheme.
However, both the JWM and WVA techniques do have their own technical advantages
in the presence of technical imperfections, e.g.,
systematic errors, misalignment errors, and limitation of detector saturation.
In particular, we revealed that, in general,
the metrological precision of the JWM scheme
cannot reach that indicated by the total FI
encoded in the PSA and PSR output data,
despite that in this scheme all the data are collected without discarding.
We also analyzed the effect of a few types of technical noise, showing that
the technical noise cannot be removed by the subtracting procedure
in the JWM scheme, but yet can be avoided or even utilized
by performing imaginary-WVs measurement.

Based on Eqs.\ (\ref{diff-P-1}), (\ref{xbar-1}),
(\ref{SV-1})-(\ref{SV-3}) and (\ref{SNR-JWM}),
we know that the amplification principle of the JWM scheme
is quite different from the WVA.
The amplification of the JWM technique
is from a statistical trick, which is classical in essence.
In contrast, the amplification of the WVA technique
is rooted in a quantum interference effect \cite{Li16}.
Indeed, the effect of WVA becomes less prominent
with the increase of measurement strength
and should be impossible in classical systems.
In contrast, the JWM technique seems applicable
to classical precision metrology
and holds similar technical advantages in the presence of such as
systematic errors, misalignment errors, and limitation of detector saturation.
An initial study can be the model proposed in Ref.\ \cite{FC14b},
where classical coin toss
was assumed to generate the WVA-like response,
which was yet negated later after more careful analysis \cite{Li16}.
We believe that the amplification effect of the JWM
can be realized in this classical model.

\vspace{0.7cm}
{\flushleft\it Acknowledgements.}---
This work was supported by the
National Key Research and Development Program of China
(No.\ 2017YFA0303304) and the NNSF of China (Nos.\ 11675016, 11974011 \& 61905174).

\appendix

\section{Real and Imaginary Weak-Value measurements}

Taking the Stern-Gerlach setup model considered in the main text,
under the AAV limit, the post-selected meter state
is obtained from the entire state of the system-plus-meter
(which evolves under the measurement coupling interaction) as
\bea
|\Phi_f\ra &\sim & \la f| e^{-idPA}(|i\ra |\Phi\ra)  \nl
&\simeq& \la f|(1-idPA)(|i\ra |\Phi\ra)  \nl
&=& \la f|i\ra (1-id A^f_w P)|\Phi\ra   \nl
&\simeq& \la f|i\ra  e^{-id A^f_w P}|\Phi\ra   \,,
\eea
where $A^f_w=\frac{\la f|A|i\ra}{\la f|i\ra}$ is the AAV WV.
Here and in the following, using ``$\sim$" means that the state is not normalized.
Then, in $x$ basis, this state can be expressed as
\bea
|\Phi_f\ra \sim
\int dx \, \Phi(x-d A^f_w) \, |x\ra  \,,
\eea
while in $p$ basis, it reads as
\bea
|\Phi_f\ra \sim
\int dp \, \Phi(p+i \frac{d}{2\sigma^2}A^f_w)\, |p\ra   \,.
\eea
Here we have used the explicit forms of the meter's initial wavefunction
$\Phi(x)=(2\pi\sigma^2)^{-1/4} e^{-x^2/4\sigma^2}$
in the $x$ basis
and $\Phi(p)=(\frac{\pi}{2\sigma^2})^{-1/4} e^{-\sigma^2 p^2}$ in the $p$ basis.
Accordingly, if performing measurement on the meter state in the $x$ basis,
the post-selected average is
$\la x\ra_f=d\, {\rm Re} A^f_w$,
while the average $\la p\ra_f=(\frac{d}{2\sigma^2}){\rm Im} A^f_w$
is obtained by measurement in the $p$ basis.
Actually the two averages are the respective centers of the wavepackets
$|\Phi(x-d A^f_w)|^2$ and $|\Phi(p+i \frac{d}{2\sigma^2}A^f_w)|^2$.
We see then that the two post-selected averages are associated, respectively,
with the real and imaginary parts of the AAV's WV.
This is the reason that in literature they are termed as
real and imaginary WV measurements.


\end{CJK*}
\end{document}